\documentclass[manuscript]{aastex61}
\usepackage{graphicx}
\usepackage{natbib}
\usepackage{amsmath}
\usepackage{hyperref}
\usepackage{bm}

\newcommand{\2}{$_2$}

\shorttitle{}

\begin{document}

\title{A New Line-By-Line General Circulation Model for Simulations of Diverse Planetary Atmospheres: initial validation and application to the exoplanet GJ 1132b}



\author[0000-0001-7758-4110]{Feng Ding}
\affil{School of Engineering and Applied Sciences, Harvard University, Cambridge, MA 02138, USA}
\email{fengding@g.harvard.edu}

\author[0000-0003-1127-8334]{Robin D. Wordsworth}
\affil{School of Engineering and Applied Sciences, Harvard University, Cambridge, MA 02138, USA}
\affil{Department of Earth and Planetary Sciences, Harvard University, Cambridge, MA 02138, USA}


\begin{abstract} 

Exploring diverse planetary atmospheres requires modeling tools that are both accurate and flexible. Here, we develop a three-dimensional general circulation model (3D GCM) that for the first time uses a line-by-line approach to describe the radiative transfer. We validate our GCM by comparing with published results done by different 1D and 3D models. To demonstrate the versatility of the model, we apply the GCM to the hot Earth-sized exoplanet GJ 1132b and study  its climate and circulation assuming an atmosphere dominated by abiotic oxygen (O\2). Our simulations show that a minor CO\2 composition can change the circulation pattern substantially, intensifying the equatorial superrotation in particular. Computation of the phase-resolved spectroscopy indicates that the vertical profile of the superrotating jet {could} be inferred in future \replaced{spectro-photometry}{spectro-photometric}  observations by the phase shift of the hotspot in the CO\2 principle absorption band centered at 667 cm$^{-1}$. 
\added{We also show that atmospheric mass could potentially be constrained by the phase amplitude in the O\2 vibrational fundamental band for planets with O\2-rich atmospheres, although further experimental and/or theoretical O\2-O\2 collision-induced absorption data at high temperatures is needed to confirm this.}
More physical schemes such as moist dynamics will be implemented in the GCM in the future so that it can be used to tackle a wide variety of planetary climate problems.

\end{abstract}

%
%



%
%
 
\section{Introduction}\label{sec:intro}

On terrestrial-mass planets, the atmospheric mass and bulk composition is expected to be extremely diverse, because the external boundary conditions (particularly exchange with the planetary interior and escape to space) dominate the atmospheric evolution  \citep{pierrehumbert2010principle, catling2017evolution}. The variety of planetary atmospheres in the present-day solar system is one example; the inferred evolution of the Earth's climate over geological timescales is another \citep{holland2014treatise, catling2017evolution}. For rocky planets outside the solar system, atmospheric characterization will be possible in the foreseeable future and great diversity is also expected \citep{birkby2018handbook, roberge2018handbook, madhusudhan2018handbook}.  Investigating these diverse planetary climates requires various tools, including flexible and accurate three-dimensional general circulation models (GCMs). 

Probably the most significant obstacle to making  flexible GCMs is the radiative transfer calculation in a realistic, vertically-inhomogeneous planetary atmosphere. Numerically computation of the radiative heating rates that drives the flow motion in GCMs is enormously computationally expensive due to the non-gray nature of the atmospheric absorption.
This arises from the   large number \deleted{($\sim 10 ^{5-6}$)} of spectral lines of the absorbing gases \added{(exceeding billions of lines for some gases)}, as well as the dependence of the line absorption on  the atmospheric state. As a result, statistical parameterizations of radiative transfer have previously been used  to make long-term integration feasible \citep{stephens1984rad,goody1989ckd, lacis1991ckd}. 
The most popular of these tools,  the correlated-\textit{k} method, has been used for several years in  studies of paleoclimate and exoplanet climates by various GCMs, including but not limited to 
the Laboratoire de M\'et\'eorologie Dynamique Generic (LMDG) model  \citep{wordsworth2011gj581d, wordsworth2013mars}, 
the Substellar and Planetary Atmospheric Radiation and Circulation (SPARC)/MITgcm \citep{showman2009sparc, kataria2015wasp43b}, 
the  Extraterrestrial Community Atmosphere Model \citep{wolf2014archean, wolf2015runaway},
and the Resolving Orbital and Climate Keys of Earth and Extraterrestrial Environments with Dynamics (ROCKE-3D)  \citep{way2017rocke3d}. 
However, statistical tools have limitations , particularly when they are used outside the range for which planetary radiative transfer is well-constrained by observations. Errors and biases  can arise if the model is applied to a situation for which it was not designed. 
Just as importantly when dealing with new temperature/pressure ranges or adding more species, a pre-processing step that relies on external line-by-line softwares to calculate the absorption coefficients is required.  This is \replaced{time-consuming}{labor extensive} and requires careful choice of approximations, which \replaced{limits}{can limit} the flexibility of GCMs based on this technique.

Here we take an alternative approach to this problem and present the development of a new GCM that for the first time uses a line-by-line approach  to calculate the radiative transfer directly. The principal advantage of this type of GCM is that our direct algorithm allows high computational accuracy and versatility when varying the atmospheric state and composition. We describe the model framework, including the line-by-line approach to  calculate the radiative transfer and validation, in Section~\ref{sec:description}. As an example application, we use this GCM to simulate the possible climates on an Earth-sized exoplanet GJ~1132b and discuss the associated spectroscopic observable features in Section~\ref{sec:1132b}. Future directions of the GCM are discussed in Section~\ref{sec:conclusions}. 

\section{Model description and validation} \label{sec:description}

\subsection{1D radiative-convective model with the line-by-line radiative transfer calculation} \label{sec:lbl}
Before implementing the line-by-line radiative transfer calculation in the 3D GCM, we first constructed a one-dimensional radiative-convective (RC) model. This  model works in a similar way to the conventional radiative-convective models \cite[e.g., ][]{manabe1964thermal}, except for the spectral radiative calculation. We use 2000-8000 points in wavenumber \added{for both shortwave and longwave calculations}; as we discuss later, we have tested that further increases in the spectral resolution have insignificant effects on the spectrally integrated flux and heating rates.
Before performing the iteration towards the radiative-convective equilibrium state, the model first reads in HITRAN and/or HITEMP line  data and computes the absorption cross-section at these discrete wavenumbers  in a temperature-log~pressure table for each absorbing species. The values of the temperature-log~pressure grids are determined by the initial condition of the surface temperature and surface pressure so that the constructed table is made full use of for long-term integrations. As the model iterates and the atmospheric state changes, new absorption cross-sections are interpolated from this table to calculate the monochromatic optical thickness and the radiative fluxes.

Absorption cross-sections are calculated by first scaling line strengths from their reference values via the formula
\begin{equation}
S_{ij}(T)= S_{ij}(T_0) \frac{Q(T_0)}{Q(T)} \frac{1 - e^{-h\nu_{ij}\slash k_BT}}{1 - e^{-h\nu_{ij}\slash k_BT_0}} \frac{e^{-h\nu_{i} \slash k_BT}}{e^{-h\nu_{i} \slash k_BT_0}}.
\end{equation}
Here $i$ and $j$ denote the ground and excited states of the transition, respectively, $S_{ij}$ is the line strength, $\nu_{ij}$ is the line location in frequency unit, $\nu_i$ is the ground state frequency, $T$ is temperature, $T_0 = 296$~K is the reference temperature, $h$ is Planck constant and $k_B$ is Boltzmann constant \citep[e.g.,][]{Rothmann1998}. Line broadening due to both collisional and Doppler effects are calculated via the usual formulae \citep{Goody1995}. The line shapes are then calculated using the Voigt profile, with the Huml{\'\i}{\v{c}}ek algorithm used to compute the complex probability function efficiently \citep{Humlicek1982,Schreier1992}.

A version of this model using fixed temperature profiles was first described in \citet{schaefer2016gj1132b}. The time-varying radiative convective model was presented in \citet{wordsworth2017transient}. To validate the 1D model, we first ensured that it reproduces  analytical results when the atmosphere is gray or semi-gray. Next, for H\2O runaway greenhouse calculations, we checked that the outgoing longwave radiation (OLR) computed by our model agrees closely with published results for the Earth's atmosphere \citep{goldblatt2013runaway}. This inter-comparison is also discussed in \citet{schaefer2016gj1132b}. 

Here, for completeness we present a summary of the model setup and describe an  additional validation test we performed -- a comparison among various 1D radiative-convective models with different radiative transfer calculations. The simulated case is an Earth-like planet with 1 bar pure CO\2 atmosphere orbiting the Sun at 1 AU. The surface has a gray albedo of 0.20. Here the line absorption coefficients for CO\2 are calculated from the 2012 HITRAN line list \citep{rothman2013hitran}, and the lines are truncated at 500 cm$^{-1}$. The spectral range of the longwave calculation is from 1 cm$^{-1}$ to 5 times the Wien peak wavenumber of the Planck function at the simulated surface temperature (3000 cm$^{-1}$ in this case). 8000 points in wavenumber are used so that the wavenumber increment is roughly 0.375 cm$^{-1}$. The CO\2 collision-induced absorption was included using the GBB parameterization \citep{borysow1998co2, baranov2004co2, wordsworth2010co2}. \replaced{An eight-stream equation is}{Eight quadrature points (four upwelling and four downwelling) are} used for both shortwave and longwave radiative flux calculations. 
Currently the eight-stream radiative calculation is only performed in the pure-absorption limit. \replaced{We are able to `paint the ground blue' with the Rayleigh scattering}{For the scattering calculation, we simply impose the enhanced reflection by Rayleigh scattering at the surface and the analytical two-stream solution is used,} because the molecular absorption and Rayleigh scattering occur in well-separated spectral regions.
In situations where clouds or aerosols are present or when significant gas absorption extends into the visible, this approximation breaks down and a multiple scattering code is required. We will describe extensions of our model into the multiple scattering regime in subsequent papers.

We compare our 1D model to two others here for the same simulated case: the Clima model  developed by Kasting and collaborators that uses the correlated-\textit{k} distribution method to describe the radiative transfer \citep{kopparapu2013clima} and the linearized flux evolution (LiFE) model  recently developed by \citet{robinson2018life} that uses a line-by-line \added{, multiple scattering} approach but combines  linear flux Jacobians to rapidly adapt radiative flux profiles to changes in atmospheric state. 

\begin{figure}[ht]
  \centering
  \includegraphics[width=\columnwidth]{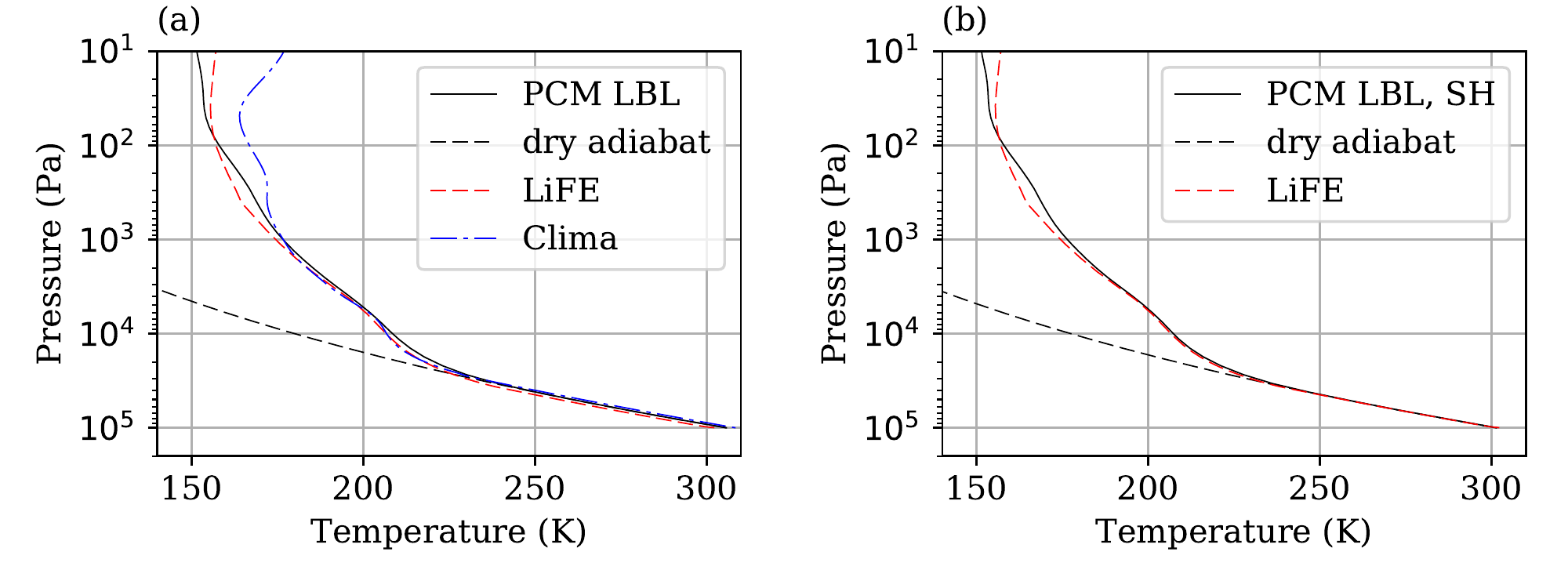}
  \caption{(a) Comparison of the equilibrium vertical temperature profiles by various models for an Earth-like planet with  1 bar pure CO\2 atmosphere (black: our 1D model, red: LiFE model, blue: Clima model).  The dashed line is the dry adiabat of CO\2 with the surface temperature of our 1D model. The vertical profiles of the Clima and LiFE model are taken from \citet{robinson2018life}. (b) Same simulation as in (a), but a  surface sensible heat flux is prescribed in our 1D model (black).}\label{fig:1dtp}
\end{figure}

\begin{figure}[ht]
  \centering
  \includegraphics[width=\columnwidth]{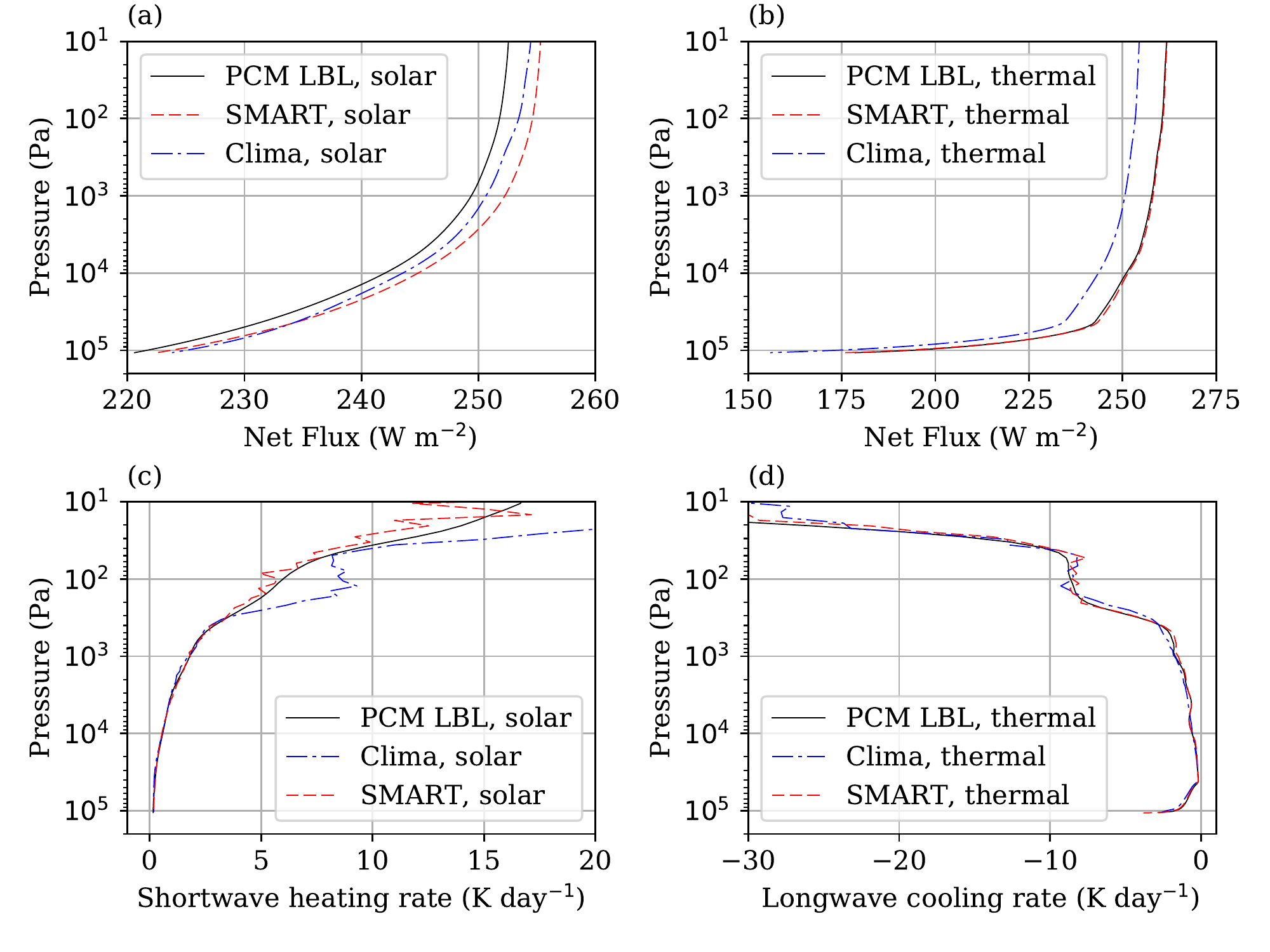}
  \caption{Vertical profiles of radiative fluxes and radiative heating rates computed with the Clima-derived equilibrium thermal structure shown in Figure~\ref{fig:1dtp}a by various models (black: our 1D model, red: SMART model, blue: Clima model):
  (a) Net solar flux.  (b) Net thermal flux. (c) Solar heating rate. (d) Thermal cooling rate. }\label{fig:fluxcomp}
\end{figure}

\begin{figure}[ht]
  \centering
  \includegraphics[width=\columnwidth]{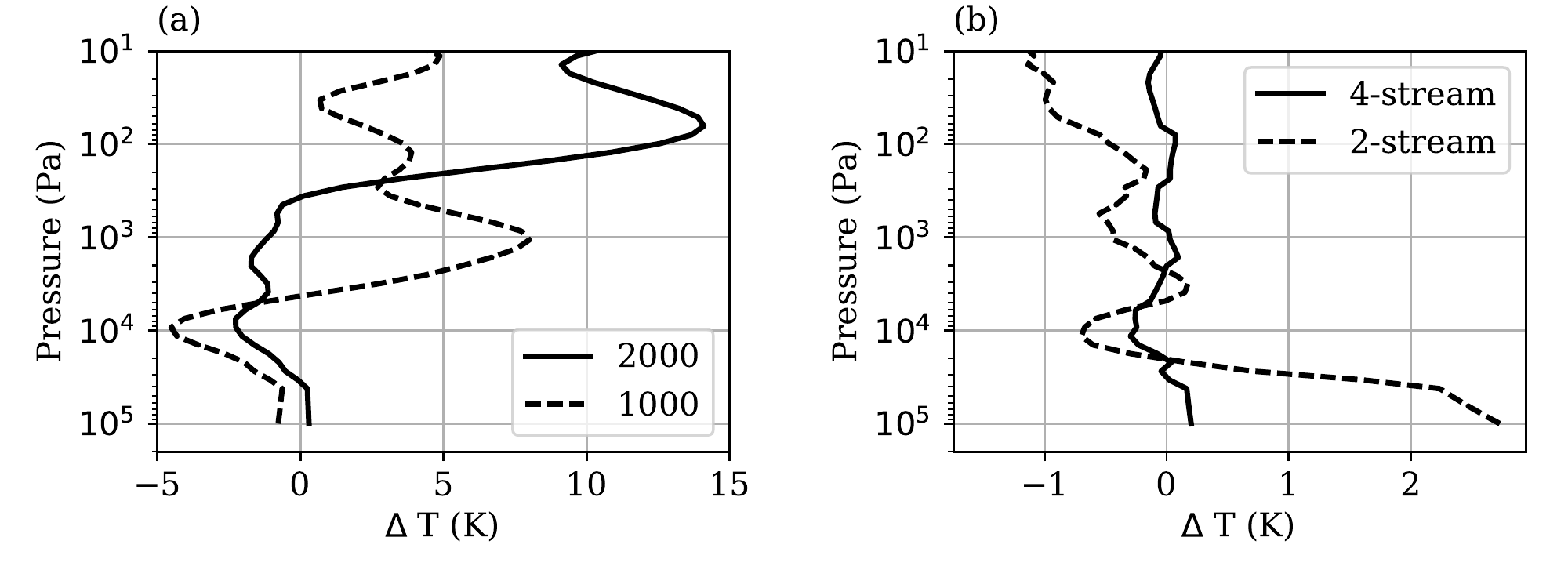}
  \caption{Difference of the temperature profiles relative to the equilibrium profile derived by our 1D model shown in Figure~\ref{fig:1dtp}a which uses 8000 spectral points and 8 quadrature points. 
  (a) Same simulation as in Figure~\ref{fig:1dtp}a, but by our 1D model with various spectral resolutions. The numbers in the legend are the number of points in wavenumber for each run. (b) Same simulation as in Figure~\ref{fig:1dtp}a, but by our 1D model with various number of quadrature points in the radiative calculation. The numbers in the legend are the number of streams  for each run.}\label{fig:nsnang}
\end{figure}

A comparison of the radiative-convective equilibrium thermal structures is shown in Figure~\ref{fig:1dtp}a. All three models yield similar surface temperatures, and hence the vertical temperature profiles in the troposphere where dry convection occurs are also similar. One notable difference is that in the upper atmosphere above 300 Pa, the two models with high spectral resolutions are consistent and both colder by $\sim 10$ K than the Clima model. \added{ In the troposphere and the lower stratosphere, our 1D model and the Clima model results are slightly warmer than that of the LiFE model, because the LiFE model has a surface boundary scheme that allows for the temperature discontinuity from the surface to the surface air, while in most conventional 1D radiative-convective models the surface temperature is taken to be  identical to the surface air temperature  for simplicity. As a test, we also prescribe a surface sensible heat flux in our 1D model. When we do this, the resulting equilibrium profile  is more consistent with the LiFE model in the troposphere and lower stratosphere (Figure~\ref{fig:1dtp}b ). }

\added{As an additional test, we also  compare the radiative fluxes directly computed using the radiative transfer routines adopted by these 1D models. Figure~\ref{fig:fluxcomp} compares the net solar and thermal flux profiles and the associated radiative heating rates for the same vertical temperature profile -- the equilibrium thermal structure determined by the Clima model. The radiative transfer routine used by the LiFE model is the Spectral Mapping Atmospheric Radiative Transfer (SMART) model \citep[developed by D.~Crisp, see][]{meadows1996venus}, which is a 1D multiple-scattering, line-by-line radiative transfer model. Our radiative calculation agrees very well with the SMART model  in the thermal infrared (IR). 
In the shortwave, the absorption of  the atmosphere in the two models are also very similar, as shown by the slope of the net shortwave flux in Figure~\ref{fig:fluxcomp}a and  the shortwave heating rate in  Figure~\ref{fig:fluxcomp}c. Our 1D model systematically underestimates the net shortwave flux by a small amount probably due to our simplification of Rayleigh scattering mentioned previously. 
In addition, while the Clima model  systematically underestimates the net thermal radiative flux by 6 W m$^{-2}$ above the tropopause (Figure~\ref{fig:fluxcomp}b), the longwave cooling rate profile calculated by the Clima model is very similar to the other two models, as shown in Figure~\ref{fig:fluxcomp}d. It is the strong near-IR absorption above 300 hPa that warms the upper atmosphere in the Clima model and even results in a  temperature inversion above 30 hPa (Figure~\ref{fig:1dtp}a). The reason for this discrepancy is unclear, but the close agreement of the two line-by-line codes suggests it is an artifact of the less accurate correlated-k approach taken in CLIMA.
}
  

Unlike  the LiFE and Clima models, which both compute the averaged radiative quantities over spectral intervals \added{\footnote{The averaged radiative quantities are derived based on fully line-by-line calculations, and the LiFE model has a much finer spectral resolution (5 cm$^{-1}$ in the thermal infrared) than the Clima model.}}, our algorithm is more like a quasi-random sampling of the absorption lines, because we perform the radiative calculation at discrete regular wavenumbers, and  the absorption line locations are extremely irregular (particularly for H\2O). 
We show the  dependence of our results on spectral resolution in Figure~\ref{fig:nsnang}a. Interestingly,  it is not necessary to use a very high spectral resolution in this radiative calculation.
First, the surface temperature is very insensitive to the spectral resolution if the resolution is high enough. The surface is only warmed by 0.8 K when the number of spectral points increases from 1000 to 8000. For the upper atmosphere, which is dominated by  narrow spectral bands with strong lines, a relatively high spectral resolution is necessary. For example, given 2000 spectral points, the temperature profile has relatively converged to the results with higher spectral resolutions  from 300 Pa to the ground, but  above 300 Pa the cooling rate may be underestimated. Slightly more efficient results could probably be obtained using an irregular spectral grid, which is something we plan to investigate in future. However, the convergence of the results in Figure~\ref{fig:nsnang}a at 8000 points indicates the general robustness of our approach even using a regular spectral grid.

\added{Another factor that affects the radiative calculation is the number of quadrature points used to compute the radiative fluxes. Most GCMs use two-stream approximations since it is computationally efficient.  In our 1D simulation, the equilibrium surface temperature rises by 2.8 K using the two-stream approach.
Figure~\ref{fig:nsnang}b shows that the lower atmosphere is warmed by the same degree using the two-stream approach, while a  four-stream approach is sufficient in this 1 bar CO\2 atmosphere simulation.
}

\subsection{Three-dimensional planetary climate model (FMS PCM)} \label{sec:fmspcm}

Our 3D GCM (hereafter referred to as FMS PCM) is developed based on the ``vertically Lagrangian'' finite-volume dynamical core \citep{lin2004fvcore} of the Geophysical Fluid Dynamics Laboratory (GFDL) Flexible Modeling System (FMS\footnote{\url{https://www.gfdl.noaa.gov/fms/}}), which solves the atmospheric primitive equations  in  spherical coordinates. The model also includes a cubed-sphere gridding technique for atmospheric dynamics \citep{ronchi1996cubed, putman2007cubed}, which improves both the computational performance and accuracy compared to  conventional latitudinal-longitudinal gridding. For example,  the cubed-sphere dynamical core allows more processors to be used for parallel computation, given the same horizontal resolution, than  spectral dynamical cores in which   the prognostic variables are decomposed by spherical harmonics and domain decomposition along the  circles of latitude for parallel computation is not allowed. 
This makes the implementation of the line-by-line radiative calculation in the dynamical core feasible given modern computational resources.

The FMS dynamical core only deals with the atmospheric transport, so to build a planetary climate model, additional physical schemes are required. The schemes we implement in FMS PCM are listed as follows:

\begin{itemize}
\item Our line-by-line radiative calculation, which was introduced in Section~\ref{sec:lbl}.
\item A convection scheme. Our scheme is similar in spirit to that described in \citet{manabe1964thermal}. The lapse rate in the unstable layer is adjusted towards a neutral state while conserving enthalpy. Currently the model only has a dry version, with no phase transitions of condensable
substances or associated cloud formation.
\item  A planetary boundary layer (PBL) scheme. The surface momentum and buoyancy fluxes are computed by the Monin-Obukhov similarity theory from the lowest level winds, temperatures, and tracer mixing ratios. A fully backward time-step method is employed for the vertical diffusion calculation in the planetary boundary layer. The diffusivity is set by the K-profile scheme \citep{troen1986pbl}.
\end{itemize}

To validate this 3D model, we use it to simulate the atmospheric collapse on synchronously-rotating rocky planets.  Our motivation for studying this setup is  the simplicity of the climate dynamics -- a single-component atmosphere dominated by an axisymmetric overturning circulation. 
We chose not to simulate the Earth's present-day climate because it is a much more complicated system. 
The specific case we have chosen is   CO\2 atmosphere collapse on an Earth-like planet orbiting a M-dwarf \citep{joshi1997tidallylocked, wordsworth2015heat}. The detailed parameters of our simulation are given by Table~1 of \citet{wordsworth2015heat}. In this simulation, the horizontal resolution is C48 ($48\times48\times6=13824$ grid boxes in total; similar to a $2^\circ \times 2^\circ$ Lat-Lon grid); the physical schemes are performed over 26 vertical layers in the hybrid sigma-pressure coordinate. 
The top pressure level is 220~Pa. As shown in Figure~\ref{fig:nsnang}a, 2000 points in wavenumber is enough for the radiative calculation for our setup. The equilibrium temperature profile in 1D simulations has converged to the profiles derived from higher spectral resolution runs. Therefore we reduce the spectral resolution to 2000 points in wavenumber \deleted{and use two-stream equations} to save computation time. \added{For this GCM run, we choose the two-stream approach to compute the radiative fluxes because the two-stream approach was adopted by the other GCM we will compare with. }
The radiative heating rate in the GCM is updated every four model hours. With these parameters, it takes about one hour for a 200-day integration using 384 CPU cores on our cluster. The results we present here are averages over the last 500 days of a 2000-day integration. 

\begin{figure}[ht]
  \centering
  \vspace{-10pt}
  \includegraphics[width=\columnwidth]{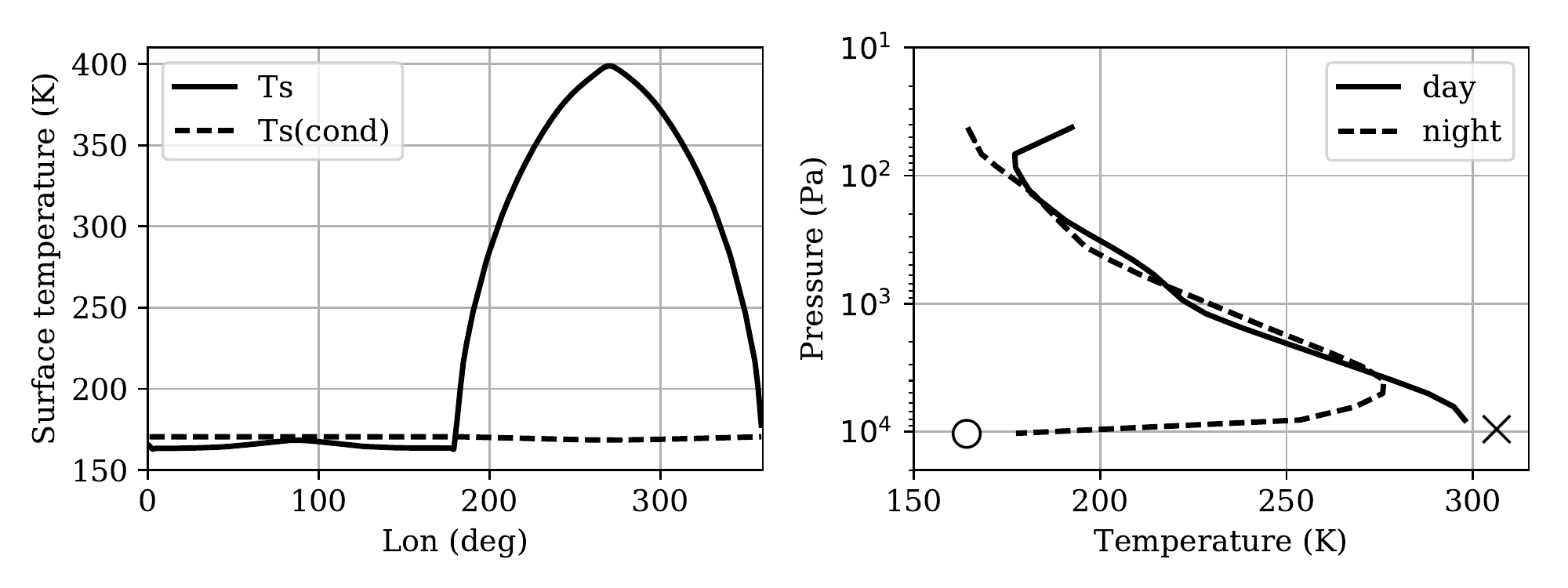}
  \caption{Left panel: equatorial surface temperature distribution (solid) and  condensation temperature corresponding to the surface pressure (dashed) computed using the FMS PCM for a synchronously-rotating rocky planet with a pure CO\2 atmosphere. The substellar point is 270$^\circ$. Right panel: hemispheric-averaged dayside (solid) and nightside (dashed) vertical temperature profiles in the same simulation. The cross and circle indicate the day and nightside surface temperatures, respectively. }\label{fig:lmd_comp}
\end{figure}

The total energy conservation of a GCM is essential for climate simulations. A simple way to check it is to calculate the global-mean top-of-atmosphere (TOA) net radiative flux. In this run, the global-mean outgoing longwave radiative flux (OLR) is 280.28~W\,m$^{-2}$ and the outgoing shortwave radiative flux (OSR) is 61.48~W\,m$^{-2}$. Given that the global-mean insolation is the same as present Earth (341.77~W\,m$^{-2}$), net TOA radiative flux is $\sim 0.01~$W\,m$^{-2}$. This also confirms that all of the physical schemes we have implemented in the GCM conserve energy very well. 
Our simulation  agrees closely with the published result simulated by  the Laboratoire de M\'et\'eorologie Dynamique (LMD) Generic Model \citep{wordsworth2015heat}: the pure CO\2 atmosphere collapses on the nightside surface when the surface pressure is less than 0.1 bar. Despite the completely different components in the FMS PCM and LMD model, from the dynamical core to the grid-box physical schemes (summarized in Table~\ref{tab:3d}), the FMS PCM produces very similar surface temperature distributions and vertical air temperature profiles (compare Figure~\ref{fig:lmd_comp} with Figure~9 in \citet{wordsworth2015heat}).

\begin{deluxetable}{lll}
\tablecaption{Comparison of GCM components between FMS PCM and LMD. \label{tab:3d}}
\tablehead{\nocolhead{GCM} & \colhead{FMS PCM} & \colhead{LMD} 
}
\startdata
Dynamical core & finite volume & finite difference \\
Radiation & line-by-line & correlated-\textit{k} distribution \\
PBL diffusion & K-profile & Mellor-Yamada 2.5 closure \\
\enddata
\end{deluxetable}

\begin{figure}[ht]
  \centering
  \vspace{-10pt}
  \includegraphics[width=\columnwidth]{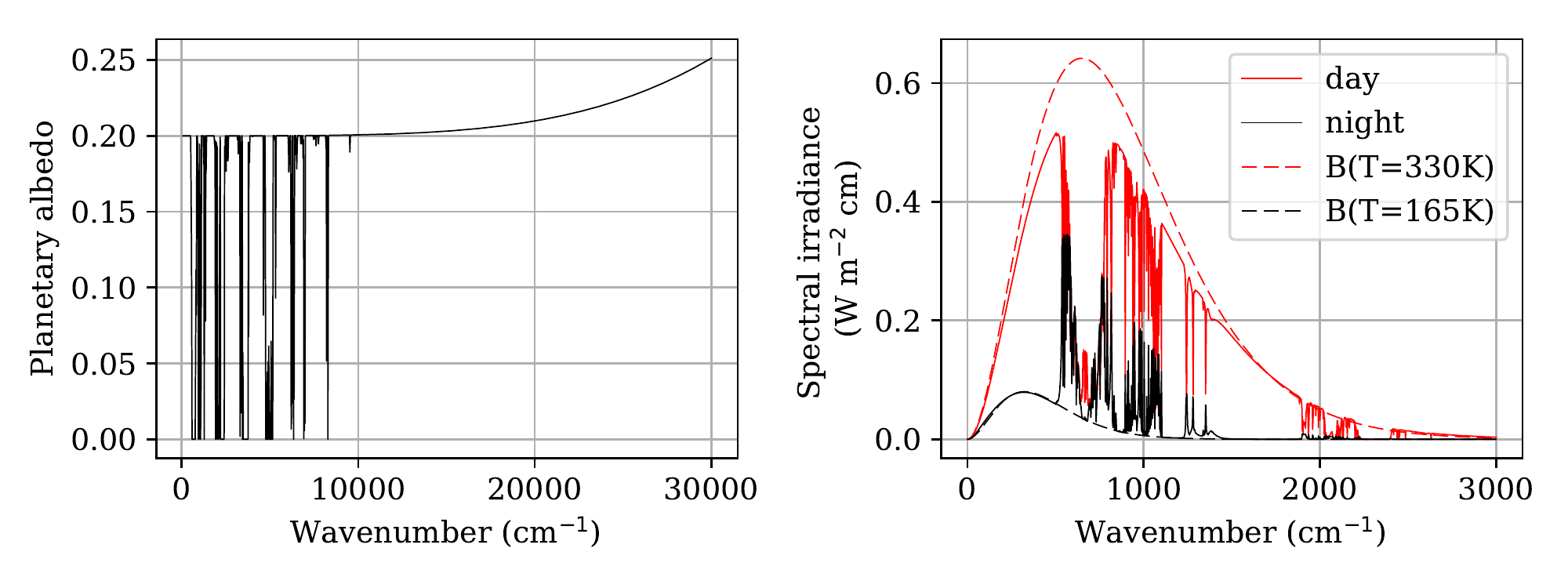}
  \caption{Left panel:  spectral distribution of the global-mean planetary albedo in the same simulation shown in Figure~\ref{fig:lmd_comp}. The surface has a gray albedo of 0.20.  Right panel: spectral flux in the thermal infrared averaged over the dayside (red solid) and the nightside (black solid). Blackbody emissions are plotted as a reference, corresponding to temperatures at 330 K (red dashed) and 165 K (black dashed). The two reference temperatures are close to the hemispheric-averaged surface temperatures. }\label{fig:osr_olr}
\end{figure}

As discussed in Section~\ref{sec:intro}, broadband averaged parameterizations in radiative transfer calculations allow relatively efficient computation in conventional GCMs. But such models fail to produce  flux data with sufficiently high resolution to be  useful for interpretation of observations. Usually, offline radiative calculations are required to get the disk-integrated spectrum \citep[e.g., ][]{robinson2011epoxi}. The FMS PCM provides another approach to calculate the disk-integrated spectrum because a relatively high-resolution radiative calculation is performed along with the model iteration. As an example, Figure~\ref{fig:osr_olr} shows the spectrally resolved planetary albedo and outgoing longwave radiation (spectral irradiance)  in this CO\2 atmosphere collapse simulation. The left panel of Figure~\ref{fig:osr_olr} is the spectral distribution of the global-mean planetary albedo. The Rayleigh scattering by the CO\2 atmosphere becomes significant for wavenumbers higher than 15000 cm$^{-1}$ and is well separated from the CO\2 absorption region, which validates our simple treatment of atmospheric scattering.  
The right panel of  Figure~\ref{fig:osr_olr} shows the spectral flux in the thermal infrared  emitted from the dayside and nightside. The IR spectroscopic feature contains very useful information regarding the vertical temperature profiles and is consistent with the profiles shown in Figure~\ref{fig:lmd_comp}. On the dayside, the air temperature decreases monotonically with height. As a result, absorption troughs emerge in the spectral region where CO\2 absorbs the thermal radiation from the lower atmosphere and re-emits with a colder temperature. On the nightside the absorption troughs become emission peaks because of the strong temperature inversion of the lower atmosphere. In particular, the big absorption trough in the CO\2 principal band centered at 667 cm$^{-1}$ becomes two emission peaks due to the strong CO\2 absorption near 667 cm$^{-1}$. The two peak emissions on both sides of 667 cm$^{-1}$ correspond to the temperature maximum at $\sim 4000$ Pa on the nightside shown in Figure~\ref{fig:lmd_comp}.
Note that there is also a clear double-peak feature near 1250 cm$^{-1}$ that is formed in a different way. It is  related to the double peak of the CO\2 collision-induced absorption feature \citep{baranov2004co2}. The CO\2 absorption in this spectral region is much weaker than that near 667 cm$^{-1}$.  

\section{Example application of FMS PCM: possible climates on an oxidized GJ~1132b} \label{sec:1132b}

In this section, we provide one demonstration of FMS PCM and use it to simulate the climates on a nearby exoplanet GJ 1132b, which was recently discovered by the MEarth ground-based transiting planet survey \citep{thompson2015gj1132b}. This rocky planet has many Earth-like parameters, including the radius, surface gravity, composition, and the spin rate assuming synchronous rotation. One major difference is that GJ 1132b receives $\sim$ 19 times more stellar insolation than the Earth and 10 times more than Venus.    \citet{schaefer2016gj1132b} discussed the possibility of abiotic O\2 buildup by the photolysis of water vapor and the subsequent atmospheric escape and mantle absorption during the magma ocean stage. The most common outcome from their simulations are tenuous atmospheres with at most a few bars of O\2 and little to no steam remaining. The existence of carbon and nitrogen-bearing species is possible but not included by the chemistry model in \citet{schaefer2016gj1132b}. The FMS PCM can be used to explore the possible climates on GJ 1132b with various surface pressures and atmospheric compositions. As a  demonstration, we consider two simple scenarios: one with 1~bar pure O\2 atmosphere; the other with an additional 1\% CO\2. Other parameters used in the GCM simulations are listed in Table~\ref{tab:1132b}.

\begin{deluxetable}{ll}
\tablecaption{Parameters used in the GCM simulations on GJ~1132b. \label{tab:1132b}}
\tablehead{ \colhead{Parameter} & \colhead{Value} 
}
\startdata
\tablenotemark{a} Planet mass $M (M_\oplus)$ & 1.62 \\
\tablenotemark{a} Planet radius $r (r_\oplus)$ &  1.16 \\
\tablenotemark{a} Surface gravity $g (g_\oplus)$  &  1.20 \\
\tablenotemark{a} Orbital period $P$ (days)   &  1.63 \\
\tablenotemark{a} Stellar flux $F (1367$ W\,m$^{-2}$) &  18.64 \\
\tablenotemark{b} Stellar spectrum & AD Leo \\
Orbital eccentricity & 0.0 \\
Obliquity & 0.0 \\
Surface roughness height (m) & $5 \times 10^{-2}$ \\
Surface albedo & 0.2 \\
Experiment A & 1 bar O\2 \\
Experiment B & 1 bar O\2 + 0.01 bar CO\2 \\
\enddata
\tablenotetext{a}{ Data taken from \citet{thompson2015gj1132b}. }
\tablenotetext{b}{ Data taken from \citet{segura2005adleo}. }
\end{deluxetable}

\subsection{Temperature and wind fields} \label{sec:1132b_t}

\begin{figure}[ht]
  \centering
  \vspace{-10pt}
  \includegraphics[width=\columnwidth]{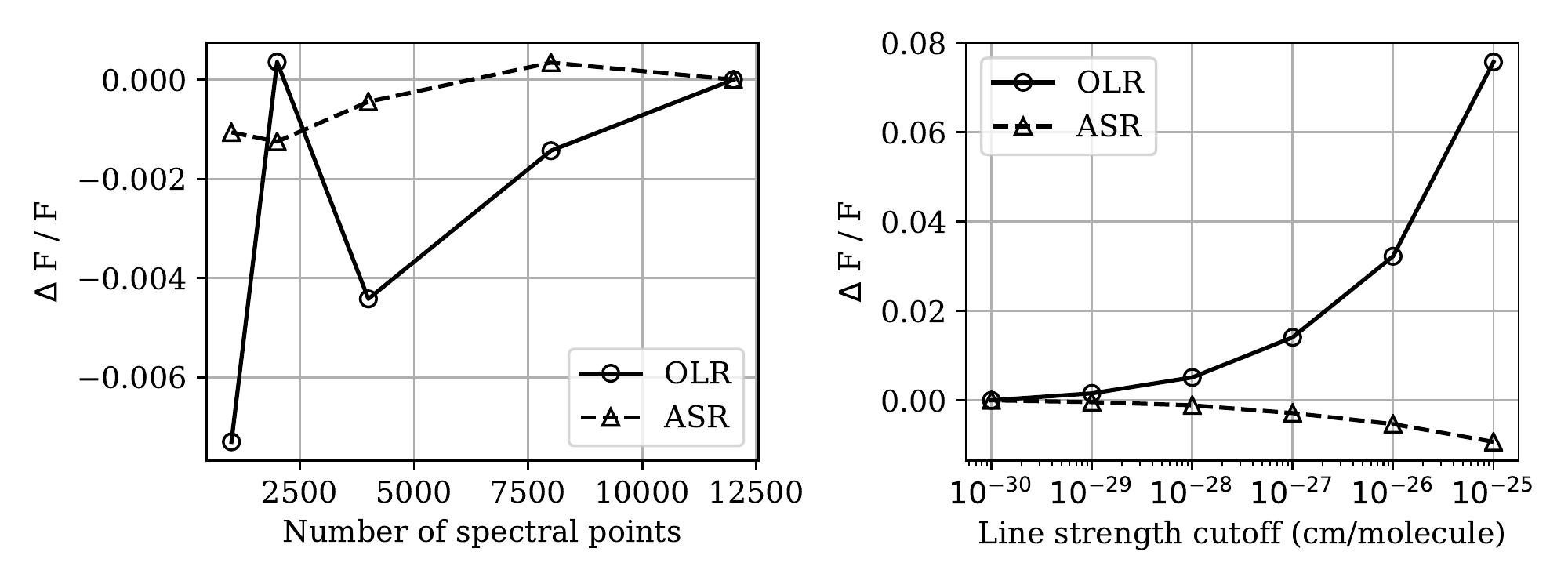}
  \caption{Relative change of radiative fluxes as a function of: (Left) number of spectral points; (Right) CO\2 line strength cutoff. The solid line with circle symbol and the dashed line with triangle symbol correspond to the outgoing longwave radiation (OLR) and the absorbed stellar radiation (ASR), respectively.  The relative change of radiative fluxes are calculated with respected to the flux with the highest resolution. }\label{fig:1132b_sens}
\end{figure}

Before running any FMS PCM simulation, we need to choose the appropriate parameters for radiative transfer calculations in the GCM. Figure~\ref{fig:1132b_sens} shows some sensitivity tests. Similar to the CO\2 atmosphere collapse simulation discussed in Section~\ref{sec:fmspcm}, it is also not necessary to use a very high spectral resolution. For this specific case, even 1000 points in both the OLR and ASR calculation yields an accuracy better than 1\%, which is good enough for long-term GCM integrations. 
Absorption by weak lines tends to be very important in this relatively hot atmosphere. We include all CO\2 lines in the GCM runs (the minimal line strength in the HITRAN database is 10$^{-30}$ cm$^{-1}$/(molecule cm$^{-2}$)). If a line strength cutoff of 10$^{-27}$ cm$^{-1}$/(molecule cm$^{-2}$) is applied, the OLR flux will increase by $\sim$1.5\%.

Our  simulation setup is basically the same as the CO\2 atmosphere collapse simulation in Section~\ref{sec:fmspcm}. The spectral resolution is also 2000 points in wavenumber, which is an appropriate choice to maintain both computation accuracy and efficiency. 
\added{However, we switch to the four-stream approach for these GCM simulations, which is  more accurate than the two-stream approach,  especially in the lower atmosphere (Figure~\ref{fig:nsnang}b). }
The O\2-O\2 collision-induced absorption (CIA) data is calculated from the HITRAN database. The transition bands and spectral ranges are summarized in Table~1 of \citet{richard2012cia}. Note that for most of the bands for this dataset, the temperature range is around  room temperature (296 K), while the equilibrium temperature ($T_{eq}$) of GJ 1132b is 548 K  assuming a planetary albedo of 20\%. The binary absorption coefficient does not vary linearly with temperature so extrapolation is  not a good choice. 
Thus in order to  simulate  radiative transfer  more accurately in a O\2-rich atmosphere on such hot planet, further experimental and/or theoretical data at high temperatures is needed. 
Here, we simply use the binary absorption coefficients corresponding to their highest temperatures. For this simulation, we also run the model by 2000 days and present an average result over the last 500 days.

\begin{figure}[ht]
  \centering
  \vspace{-10pt}
  \includegraphics[width=\columnwidth]{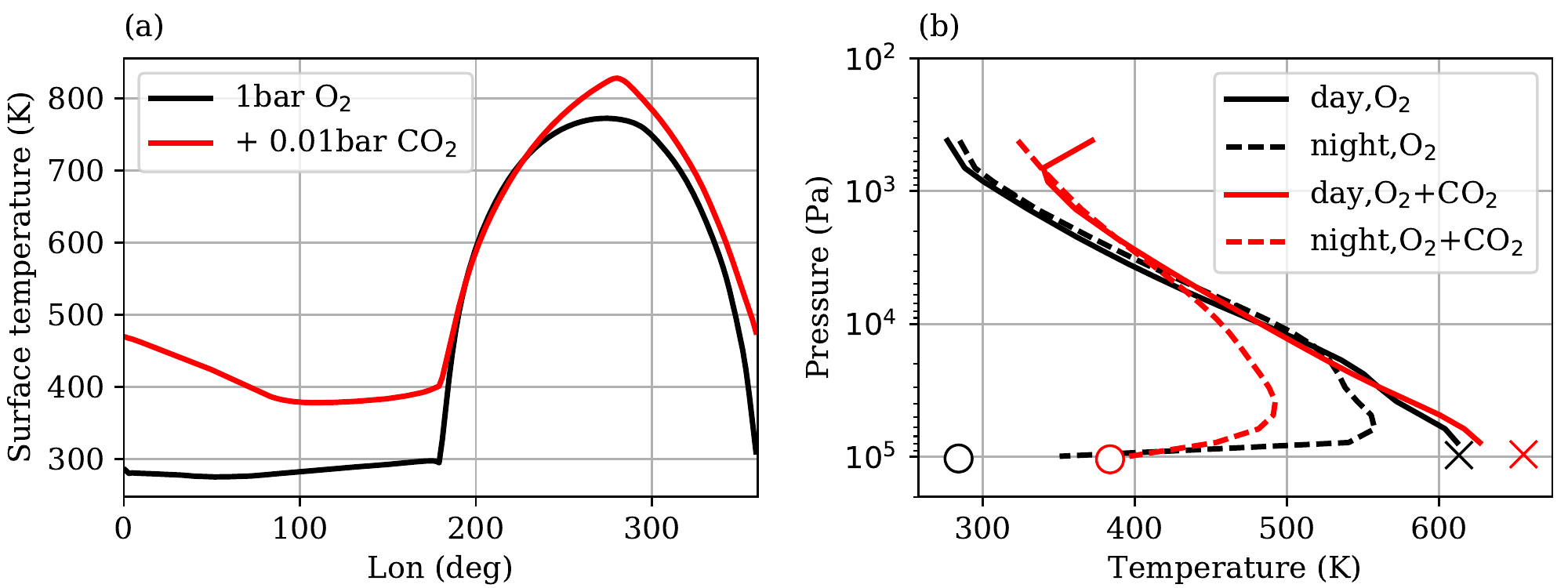}
  \caption{(a) Equatorial surface temperature distribution in the pure O\2 (black) and the O\2+CO\2 (red) simulations. The substellar point is 270$^\circ$.  (b) Hemispheric-averaged dayside (solid) and nightside (dashed) vertical temperature profiles in the pure O\2 (black) and the O\2+CO\2 (red) simulations. The  day and nightside surface temperatures are marked by crosses and circles, respectively. }\label{fig:1132b_t}
\end{figure}

The equatorial surface temperature distributions and vertical temperature profiles in these two experiments are shown in Figure~\ref{fig:1132b_t}. While the peak surface temperatures are much higher, the  general structure here is analogous to our CO\2 collapse simulation shown in Figure~\ref{fig:lmd_comp}, including the larger variations of surface temperatures on the dayside and the strong air temperature inversions in the lower atmosphere on the nightside. For the pure O\2 simulation, the upper atmosphere is much colder than the skin temperature ($T_{skin} = T_{eq} / 2^{1/4} = 461$ K), while in 1D radiative-convective simulations (results not shown) the upper atmosphere is indeed nearly isothermal with the skin temperature due to the weak absorption by O\2. Thus the atmospheric dynamics in the GCM plays a remarkable role in cooling the upper atmosphere efficiently. The energetics within a global-averaged atmospheric column can be simplified as \citep[See][Chap. 13, Eq.~(13.21)]{peixoto_physics_1992}
\begin{equation} \label{eq:energy}
\frac{\partial }{\partial p} \left[ \omega \left(c_p T+gz+\frac{u^2+v^2}{2} \right) \right] = \dot{Q}_{dia}
\end{equation}
where $\omega = Dp/Dt$ is the vertical pressure velocity, $c_pT, gz, (u^2+v^2)/2 $ are the enthalpy, potential energy and kinetic energy of air respectively. $\dot{Q}_{dia}$ is the diabatic heating rate, which in our GCM simulation  is solely due to the radiative heating  in the upper atmosphere. Because the temperature profile in this upper atmosphere deviates significantly from the radiative-equilibrium value (i.e., the skin temperature), the radiative heating rate is non-trivial and must be balanced by the  convergence of vertical energy transport represented by the left-hand side of Eq.~\ref{eq:energy}. 

Figure~\ref{fig:1132b_t} also shows that adding 1\% CO\2 to the atmosphere raises both the surface temperature and near-surface air temperature due to the greenhouse effect of CO\2. However, the lower atmosphere on the nightside is cooled  because of the strong temperature inversion. CO\2 is only an IR absorber here whose radiative effect depends on the lapse rate of the atmosphere. In other words, the surface and lower atmosphere are more closely connected by the enhanced atmospheric emissivity. A simple one-layer atmosphere model for the nightside  helps to understand this effect. We assume the surface is a blackbody with temperature of $T_s$, and the overlying atmosphere has a temperature of $T_a$ and a gray emissivity of $\epsilon$. Ignoring the surface sensible heat exchange between the atmosphere and the surface, the energy budgets at the surface and  at the top of atmosphere are 
\begin{eqnarray}
\epsilon \sigma T_a^4 & = & \sigma T_s^4 \label{eq:esurf} \\
OLR_{n} & = & \epsilon (2-\epsilon) \sigma T_a^4 = (2-\epsilon) \sigma T_s^4 \label{eq:enight}
\end{eqnarray}
where $\sigma$ is the Stefan-Boltzmann constant, $OLR_{n}$ is the mean outgoing longwave radiation on the nightside that also equals to the heat transport from the dayside to the nightside.
{First, Eq.~\ref{eq:esurf} shows that for a gray atmosphere $(\epsilon < 1), T_a > T_s$ and the temperature inversion is necessary on the nightside.} Then we can see that an increase of the atmospheric emissivity tends to warm the surface and cool the overlying atmosphere, assuming a small change in the $OLR_{n}$ from Eq.~\ref{eq:enight} (see also Section 3.3 of \citet{wordsworth2015heat} for a similar analysis assuming a uniform atmospheric temperature). However, the change of temperature in the GCM runs is much more complicated than this simple model with these assumptions. First, more energy is transported to the nightside in the O\2+CO\2 run. Second, the surface sensible heat is comparable to the radiative fluxes. Third, the atmosphere is not gray  but only absorbs strongly in some spectral regions by CO\2. {Our qualitative explanation here does not take  these effect into account. \citet{koll2016heat} developed a two-column radiative-convective-subsiding model to study temperature structures of dry tidally-locked rocky exoplanets, which incorporates the atmospheric dynamics by constraining 
the large-scale wind speed using heat engine efficiency. Figure~10 in \citet{koll2016heat} shows similar results to our GCM simulations: an increase in the longwave optical thickness warms the nightside surface but cools the nightside overlying atmosphere, when the atmosphere is not optically thick and has small wave-to-radiative timescale ratio (for our GCM runs on GJ~1132b it is $\sim 0.04$).}

Note that GJ 1132b has an Earth-like spin rate assuming synchronous rotation. The effects of planetary rotation on the atmospheric circulation becomes significant compared to our CO\2 collapse simulation shown in Figure~\ref{fig:lmd_comp}. We can verify this by comparing the planetary radius  with the equatorial Rossby deformation radius $L_{RO}^2 \sim (\sqrt{RT_{eq}} / \Omega a) a^2$, where $R$ is the specific gas constant, $T_{eq}$ the equilibrium temperature, $\Omega$ the spin rate of the planet, $a$ the planetary radius \citep[e.g., ][]{koll2015phasecurve, pierrehumbert2016nondilute}. The non-dimensional parameter $ (\sqrt{RT_{eq}} / \Omega a)$ for GJ 1132b is 1.14. As a comparison, this parameter is 0.62 for the Earth and 18 for our CO\2 collapse simulation. As a consequence of planetary wave perturbations, the symmetry of the surface temperature around the substellar (270$^\circ$) and antistellar point (90$^\circ$) in Figure~\ref{fig:lmd_comp} breaks down in the two GJ 1132b simulations, as shown in Figure~\ref{fig:1132b_t}a. 

\begin{figure}[ht]
  \centering
  \vspace{-10pt}
  \includegraphics[width=\columnwidth]{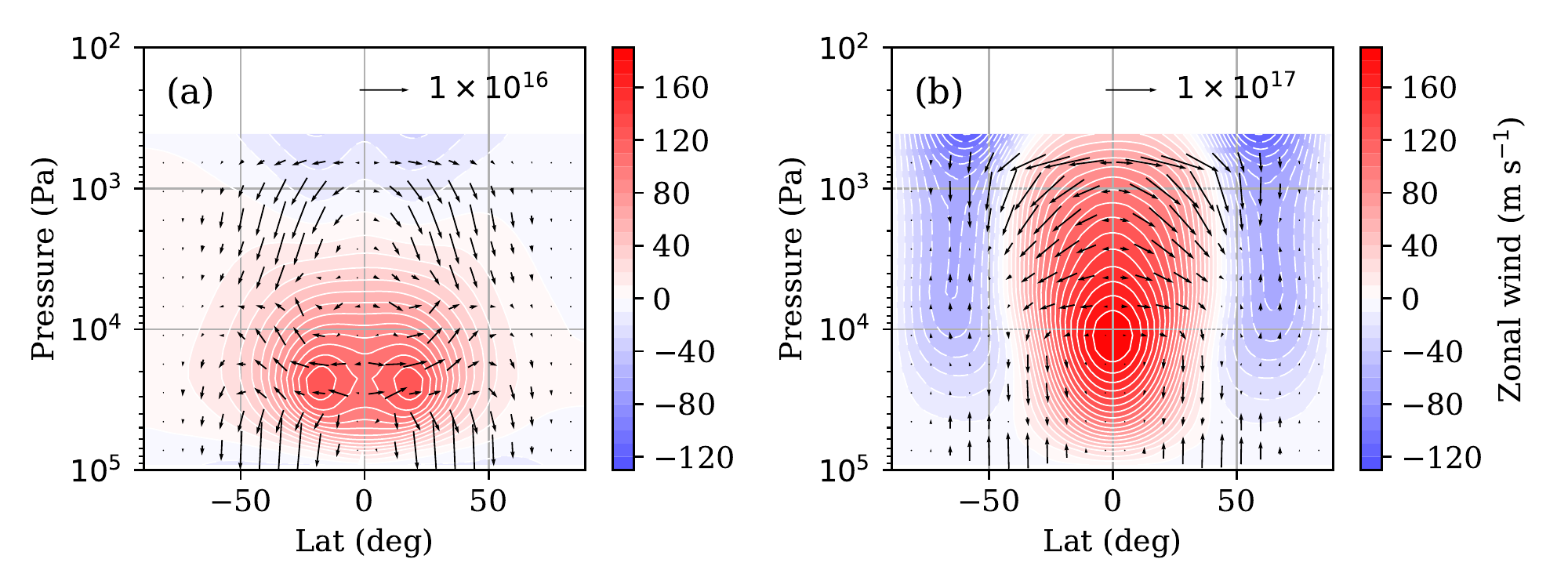}
  \caption{(a) Zonal-mean cross sections of the zonal wind component (color shading, unit: m s$^{-1}$) and the Eliassen-Palm flux (arrow vectors, unit: m$^3$ (horizontal)) for the 1 bar pure O\2 simulation. (b) Same as (a), but for the O\2+CO\2 simulation. The arrow scale for the EP flux is indicated at the upper right; note that it is different in the two panels. The vertical components of the EP flux in the two panels are rescaled by a factor of $10^{-5} / \bar{\rho}(p) $ where $\bar{\rho}(p)$ is the reference density profile.}\label{fig:1132b_epflux}
\end{figure}

Another consequence of the planetary wave perturbations is   equatorial superrotation. The zonal-mean cross sections of the zonal wind are plotted in Figure~\ref{fig:1132b_epflux}. Equatorial superrotation is a common feature of hot-Jupiter simulations with speeds of 1--4 km s$^{-1}$ \citep{showman2002gcm, dobbsdixon2008gcm, menou2009gcm}. 
It has also been observed indirectly by the eastward displacement of the hottest region from the substellar point in light curves of the hot Jupiter HD~189733b \citep{knutson2007hd189733b} and by the the ellipsoidal/beaming amplitude discrepancy on Kepler-76b \citep{faigler2013beer}. The equatorial super-rotating jets in the two experiments are of the order of 100 m s$^{-1}$, much slower than the simulated jet speed on hot-Jupiters, probably because of the efficient surface momentum dissipation and the heavier molecular weight of the O\2 atmosphere compared to the H\2-dominated atmosphere on hot-Jupiters. The effects of molecular weight on the zonal wind speed was discussed in \citet{zhang2017bulkcomposition}.
To illustrate the underlying mechanism of   equatorial superrotation, we  calculated the Eliassen-Palm (EP) flux in the latitude-pressure coordinate, which is a useful  tool for diagnostics of wave activities in the Earth's climate \citep[e.g.,][]{edmon1980epflux}. The horizontal and vertical components of the EP flux relate to the meridional eddy momentum and heat flux in the atmosphere, respectively. 
The divergence of the EP flux represents an internal forcing of the zonal jet by disturbances, so that the  equatorial superrotation in Figure~\ref{fig:1132b_epflux} can be explained by the divergence of the vectors at the equator. Comparing the two panels in Figure~\ref{fig:1132b_epflux} shows that adding 1\%  CO\2 to the atmosphere changes the circulation pattern substantially. It  amplifies the contrast of radiative-equilibrium temperature between the dayside and nightside, and therefore increases the wave amplitude and  the equatorial super-rotating jet speed. More angular momentum is transported equatorward from 
high latitudes, creating two easterly jets above 50$^\circ$. The position of the jet core also changes. In the pure O\2 simulation, two jet cores form in the subtropics associated with the meridional mean angular momentum transport, although the air above the equator still super-rotates. With an additional 1\%  CO\2, one well-formed jet core is found at 0.1 bar above the equator. 

\subsection{Thermal phase curves}
\begin{figure}[ht]
  \centering
  \vspace{-10pt}
  \includegraphics[width=\columnwidth]{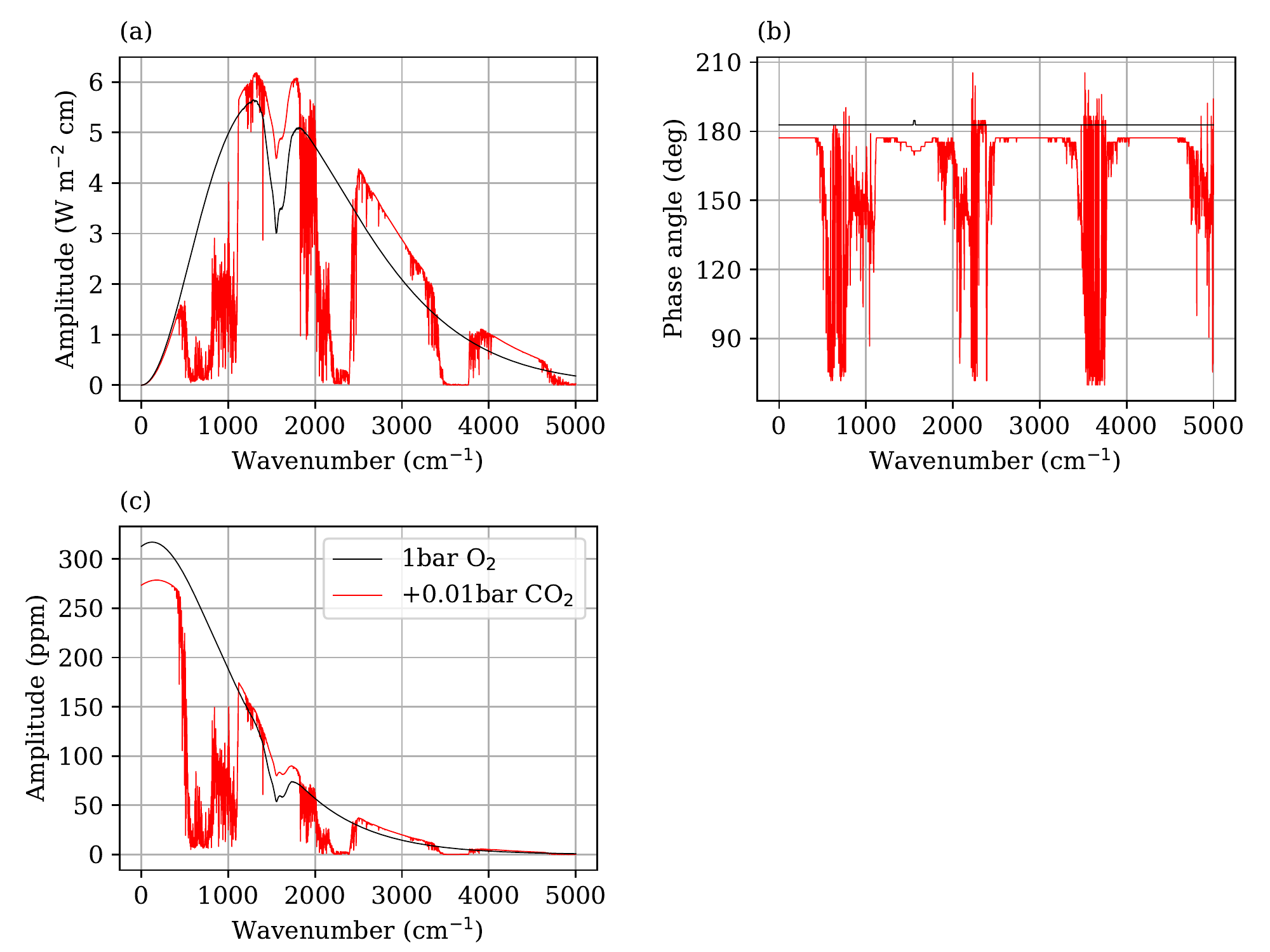}
  \caption{The spectral distribution of the thermal phase curve in the 1 bar pure O\2 (black) and the O\2+CO\2 (red) simulation: (a) the amplitude (`variation spectrum' following \citet{selsis2011phasecurve}); (b) the phase angle corresponding to the highest emission; (c) Same as (a) but taking into account the star-planet contrast. The phase amplitude is expressed in ppm. }\label{fig:1132b_phase}
\end{figure}

Thermal phase curves, which are the variations of the apparent infrared emission of an planet with its orbital phase, have been observed for hot Jupiters and used to infer their temperature distribution and atmospheric circulation  \citep[e.g.,][]{knutson2007hd189733b}. Furthermore,
\citet{selsis2011phasecurve} addressed the ability to use the multiband phase curves to characterize the  atmosphere on hot terrestrial exoplanets with next-generation observations. Here we compute the disk-integrated and observer-projected spectrum from GJ 1132b following {\citet{Cowan2008} and} \citet[][Appendix C]{koll2015phasecurve}, $F_{\nu}(\theta)$ where $\theta$ is the phase angle (we assume $\theta = 0^\circ$ at primary transit and $\theta = 180^\circ$ at secondary eclipse). Then we plot the  spectral distribution of the amplitude of $F_{\nu}(\theta)$ and the phase angle corresponding to the highest emission $\theta_{max}$ in Figure~\ref{fig:1132b_phase}a and \ref{fig:1132b_phase}b, respectively. 
{Note that  $F_{\nu}(\theta)$ here is the disk-integrated and observer-projected flux from the planet. Then the star-planet contrast seen by a distant observer should be $F_{\nu} r_p^2 / F_{\nu, \ast} r_\ast^2$, where $F_{\nu, \ast}$ is the spectral flux from the M-dwarf GJ~1132 at the same wavenumber, $r_p$ and $r_\ast$ are the radius of the planet GJ~1132b and the star GJ~1132 respectively.}  

The spectral distribution of the amplitude of $F_{\nu}(\theta)$ is called the `variation spectrum' in \citet{selsis2011phasecurve}. In our simulations, it {has a similar shape} to the emission spectrum from the dayside atmosphere (e.g., Figure~\ref{fig:osr_olr}) because the temperature contrast between the dayside and nightside atmosphere decreases with height (shown in Figure~\ref{fig:1132b_t}b) {and window regions exhibits larger variations than optically-thick spectral regions}.  \citet{meadows2017o2} performed a thorough review on   the identification of an oxygenic photosynthetic biosphere and discrimination between biological and abiotic sources of O\2, but only with a focus on  transmission spectroscopy  and direct imaging of exoplanets. 
 As described in \citet{selsis2011phasecurve}, the variation spectrum is an interesting possibility for molecular signature detections. In our simulations, O\2 absorption in the variation spectrum is  significant in its vibrational fundamental band near 1500 cm$^{-1}$. \added{Figure~\ref{fig:1132b_phase}c also shows the spectral distribution of the phase amplitude, but taking into account the star-planet contrast. The signal near 400 cm$^{-1}$ is $\sim 300$ ppm, which is slightly smaller than the the day-night thermal emission contrast of a bare rock GJ~1132b in the mid-IR \citep[373 ppm,][]{koll2016heat} because of the heat redistribution in the GCM simulation. We estimate the precision of \textit{James Webb Space Telescope (JWST)} in the mid-IR (6.6-8.8 $\mu$m broadband, F770W on MIRI) assuming photon noise following \citet{koll2016heat}.  The imperfect instrument throughput is accounted for by a factor of 1/3 \citep[Figure~3, ][]{Glasse2010}. For one full orbit integration, the $1\sigma$ uncertainty interval for the phase amplitude is 7 ppm. The O\2 absorption in this band reduces the flux ratio by $\sim 20$ ppm compared to the bare rock scenario. The value is much smaller in the O\2+CO\2 simulation, as Figure~\ref{fig:1132b_phase}c shows, which is probably associated with the enhanced near surface temperature inversion on the nightside described in Section~\ref{sec:1132b_t}. Therefore, our simulations indicate that if the O\2-dominated atmosphere is thinner than 1 bar, it might be difficult to  distinguish  from a bare rock case using the IR phase curves. 
 }

Other than the amplitude of the phase curve, our calculation indicates that the spectral distribution of the phase angle corresponding to the highest emission $\theta_{max}$ can also be used for molecular detections.  Moreover, the $\theta_{max}$ distribution provides valuable information regarding the vertical profile of the hotspot shift in the atmosphere, which is associated with atmospheric dynamics. In the 1 bar pure O\2 simulation, $\theta_{max}$ in the O\2 fundamental band indicates that \replaced{the hotspot in the lower atmosphere moves westward relative to the substellar point by $\sim 5^\circ$}{there is insignificant hotspot shift in the lower atmosphere relative to the surface}. In the O\2+CO\2 simulation, $\theta_{max}$ in the same band indicates a 6$^{\circ}$ eastward shift of the lower atmosphere hotspot, as a result of the circulation change discussed in Section~\ref{sec:1132b_t}. The absorption bands of CO\2 have more complicated features. We can take the principle absorption band centered at 667 cm$^{-1}$ as an example. In both the window region with little absorption and  the optically-thick region near the 667 cm$^{-1}$ center, $\theta_{max} \sim 180^\circ$, indicating small hotspot shifts at the surface and top of the atmosphere. But on the shoulder of this absorption band, $\theta_{max}$ can be as low as 75$^\circ$, indicating a significant eastward shift of the hotspot by 105$^\circ$ in the middle atmosphere. 

\begin{figure}[ht]
  \centering
  \vspace{-10pt}
  \includegraphics[width=\columnwidth]{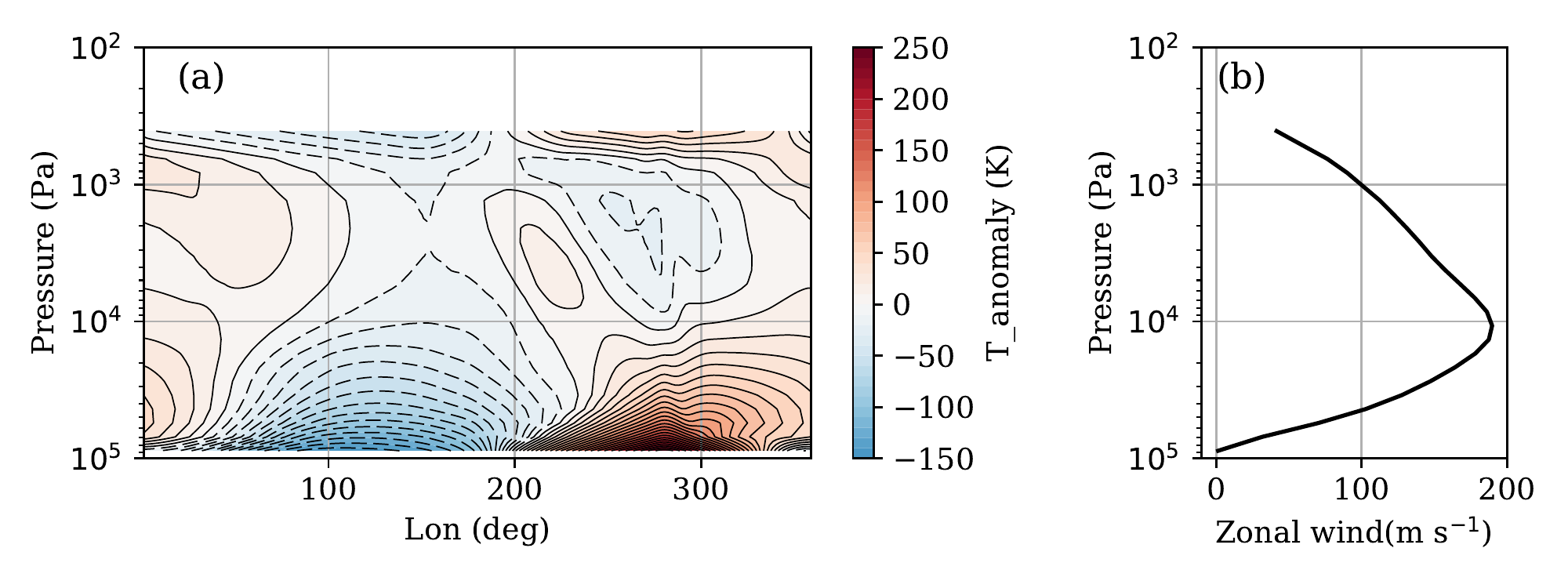}
  \caption{(a) Vertical cross section of the air temperature anomaly (obtained by subtracting the zonal mean) along the equator in the O\2+CO\2 simulation. The substellar point is 270$^\circ$. (b) Vertical profile of the zonal-mean zonal wind speed at the equator.  The longitudinal tilting of the hotspot correlates well with the super-rotating jet. 
}\label{fig:1132b_tandu}
\end{figure}

Figure~\ref{fig:1132b_tandu}a shows the vertical cross section of the air temperature anomaly (obtained by subtracting the zonal mean) along the equator  in the O\2+CO\2 simulation to confirm our inferences from  the $\theta_{max}$ distribution. There is a remarkable eastward shift of the hotspot in the middle atmosphere near 10$^4$ Pa, which is also the altitude of the jet core shown in Figure~\ref{fig:1132b_epflux}b. Moreover, the longitudinal tilting of the hotspot correlates well with this equatorial super-rotating jet, because the hotspot shift is the consequence of atmospheric dynamics. Previous studies suggested that the eastward hotspot shift is caused by the equatorial super-rotating jet \citep[e.g.,][]{zhang2017bulkcomposition, hammond2018wavemeanflow}, with  
\citet{zhang2017bulkcomposition} focusing on the heat advection by the equatorial super-rotating jet and \citet{hammond2018wavemeanflow} focusing on the Doppler-shifting stationary wave response. Note that until now theoretical work has only been done in the shallow-water framework. Further investigation is needed to explore whether baroclinicity is important for wave-mean flow interactions and therefore the longitudinal tilting of the hotspot in this situation of large horizontal temperature variation.

\section{Discussion} \label{sec:conclusions}

We have described a new 3D climate model, FMS PCM, which  uses a line-by-line approach to describe the  radiative transfer. \added{We have validated our radiation code when the absorbing species are H\2O and CO\2. At present, the same number of spectral points are used for both shortwave and longwave calculation and are evenly spaced. For other absorbing species which have narrow absorption spectral region or relatively regular line positions (e.g., CO), we may need adaptive spectral grids  and make the radiative calculation more efficient. We plan to  investigate this in future.}

Before running the GCM, some preparation is required to determine the appropriate setup (most importantly the spectral resolution) and maintain the computational accuracy and efficiency for long-term integrations. Compared to  conventional GCMs that use parametrized  radiative calculations, however, our model is designed to be both more accurate and flexible for the study of  diverse planetary atmospheres. Moreover, because all spectral quantities are calculated  within FMS PCM, this model can help interpret next-generation exoplanet observations. For example, we show that the emission spectrum in the CO\2 principle absorption band centered at 667 cm$^{-1}$ could be used to infer the near surface temperature inversion on the nightside of tidally-locked terrestrial planets. 

We have used FMS PCM 
  to  study  possible O\2-dominated atmosphere on GJ~1132b. \added{Currently the atmospheric composition of this planet is very poorly constrained, and so future studies should explore the dynamical and climate implications of a wider range of possibilities.} The simulation results show that a minor amount of CO\2 in the atmosphere can change the climate substantially, notably by enhancing the near-surface temperature inversion on the nightside and  intensifying  the equatorial superrotation and high-latitude easterlies. 
Moreover, the vertical profile of the equatorial superrotation could be inferred from the phase shift in the CO\2 principle absorption band centered at 667 cm$^{-1}$. As an example of using GCM simulations to help interpret exoplanet atmospheric observations, the metallicity of a hot-Jupiter WASP-43b was constrained in a study of  \citet{kataria2015wasp43b} by comparing a range of GCM results to the \textit{Hubble Space Telescope (HST)}/WFC3 spectrophotometric phase curve measurements. 
While the phase-resolved spectroscopic feature on GJ 1132b we have discussed here lies  in the mid-IR rather than  the visible/near-IR, it should be possible to perform such measurements using next-generation instruments, such as \textit{JWST} \citep{selsis2011phasecurve, beichman2014jwst, cowan2015jwst, koll2016heat}. {\citet{koll2016heat} estimated the amount of time that is required to measure the thermal phase curves of short-period terrestrial  planets with \textit{JWST}/MIRI,  which is similar to the time required to detect molecules via stacked transit spectroscopy.} \added{Although in our GCM simulations with 1 bar O\2 the molecular O\2 absorption is hard to be identified in its fundamental band near 1500 cm$^{-1}$ due to  small signal-to-noise (S/N) ratio, large S/N ratio is still expected for thicker O\2 atmospheres.  The broadband IR phase amplitude can be used to constrain the atmospheric mass, as suggested by \citet{koll2015phasecurve}. Note that the O\2-O\2 collision-induced absorption data in the vibrational fundamental band is only  measured below 360 K and the temperature dependence is very non-linear \citep{baranov2005o2}. Further experimental and/or theoretical data at high temperatures is required to improve the radiative calculation at high temperatures. }

So far FMS PCM is only a dry model.  Moist dynamics, including  phase transitions of  condensible substances and formation and transport of  condensates, is not yet implemented. We also plan in future to extend the radiative transfer model to allow full treatment of multiple scattering by atmospheric particulates (clouds and hazes).  Nonetheless,  the dry runs discussed in the paper demonstrate the high potential of this approach. 
We believe  that this new type of GCM model will allow the detailed  study of  a wide variety of important planetary climate problems in future, including the climate dynamics of condensible-rich atmospheres \citep{ding2016condensible, pierrehumbert2016nondilute, ding2018pure}, atmospheric characterization of exoplanets near the inner and outer edges of the habitable zone \citep{wordsworth2011gj581d, yang2013innerhz, kopparapu2016innerhz}, and water loss and abiotic oxygen build-up in water-rich atmospheres \citep{wordsworth2014abotico2}.

\acknowledgments
Support for this work was provided by  NASA grants 
 NASA NNX16AR86G and 
 80NSSC18K0829.
The GCM simulations in this paper were carried out on the Odyssey cluster supported by the FAS Division of Science, Research Computing Group at Harvard University. 
The authors would like to thank Tyler D. Robinson for insightful comments of this work and for facilitating model comparisons with the Clima and LiFE model, and 
Wanying Kang and Hannah Diamond-Lowe for helpful discussions.


\bibliography{fmspcm}
\bibliographystyle{aasjournal}
%



\listofchanges

\end{document}